\documentclass[sigconf]{acmart}
\pdfoutput=1 

\usepackage{amsmath,amssymb,amsfonts}
\usepackage{xcolor}
\usepackage{gensymb}
\usepackage{url}
\usepackage{xspace}
\usepackage{rotating}
\usepackage{array}

\usepackage{color}
\usepackage{booktabs}
\usepackage{comment}
\usepackage{multirow}
\usepackage{enumitem}
\usepackage{rotating}
\usepackage{balance}

\usepackage{array}
\usepackage{adjustbox}
\usepackage{colortbl}

\usepackage{amsmath}
\usepackage{amsfonts}
\usepackage{amssymb}
\usepackage{graphicx}
\usepackage{amsthm}
\usepackage{dsfont}
\usepackage{url}
\usepackage{hyperref}
\usepackage[colorinlistoftodos,prependcaption,textsize=footnotesize]{todonotes}

\usepackage{tikz,xstring,pgfplots,psfrag}
\usetikzlibrary{shapes,arrows}
\usetikzlibrary{calc,patterns,decorations.pathmorphing,decorations.markings}
\usetikzlibrary{positioning,automata}
\usetikzlibrary{pgfplots.groupplots}
\pgfplotsset{compat=newest}
\tikzstyle{block} = [draw, rectangle, minimum height=2em, minimum width=2em]
\tikzstyle{sum} = [draw, circle, node distance=1.5cm]
\tikzstyle{input} = [coordinate]
\tikzstyle{output} = [coordinate]
\usepgfplotslibrary{fillbetween}

\usepackage{xspace}

\usetikzlibrary{external}

\makeatletter
\g@addto@macro\normalsize{%
 \setlength\abovedisplayskip{2pt plus 2pt minus 1pt} 
 \setlength\belowdisplayskip{2pt plus 2pt minus 1pt} 
 \setlength\abovedisplayshortskip{2pt plus 2pt minus 1pt} 
 \setlength\belowdisplayshortskip{2pt plus 2pt minus 1pt} 
}
\makeatother

\makeatletter
\let\origsection\section
\renewcommand\section{\@ifstar{\starsection}{\nostarsection}}
\newcommand\nostarsection[1]
{\sectionprelude\origsection{#1}\sectionpostlude}
\newcommand\starsection[1]
{\sectionprelude\origsection*{#1}\sectionpostlude}

\newcommand\sectionprelude{%
}
\newcommand\sectionpostlude{%
}

\let\origsubsection\subsection
\renewcommand\subsection{\@ifstar{\starsubsection}{\nostarsubsection}}
\newcommand\nostarsubsection[1]
{\subsectionprelude\origsubsection{#1}\subsectionpostlude}
\newcommand\starsubsection[1]
{\subsectionprelude\origsubsection*{#1}\subsectionpostlude}

\newcommand\subsectionprelude{%
}
\newcommand\subsectionpostlude{%
}
\makeatother

\AtBeginDocument{%
  \providecommand\BibTeX{{%
    \normalfont B\kern-0.5em{\scshape i\kern-0.25em b}\kern-0.8em\TeX}}}

\newcommand{\squishlist}{ 
   \begin{list}{$\bullet$}
    { \setlength{\itemsep}{0pt}      \setlength{\parsep}{3pt} 
      \setlength{\topsep}{3pt}       \setlength{\partopsep}{0pt}
      \setlength{\leftmargin}{1.3em} \setlength{\labelwidth}{1.3em}
      \setlength{\labelsep}{0.5em} } }

\newcommand{\squishend}{
  \end{list}  }

\renewcommand{\theenumi}{\Alph{enumi}}

\usepackage{pifont}

\newcommand{\R}{\mathbb{R}}

\usepackage{amsmath}

\DeclareMathOperator*{\argmin}{arg\,min}

\newcommand{\nnskip}{\vspace{-0.15cm}}

\copyrightyear{2020}
\acmYear{2020}
\setcopyright{acmcopyright}\acmConference[SEAMS '20]{IEEE/ACM 15th International Symposium on Software Engineering for Adaptive and Self-Managing Systems}{October 7--8, 2020}{Seoul, Republic of Korea}
\acmBooktitle{IEEE/ACM 15th International Symposium on Software Engineering for Adaptive and Self-Managing Systems (SEAMS '20), October 7--8, 2020, Seoul, Republic of Korea}
\acmPrice{15.00}
\acmDOI{10.1145/3387939.3391568}
\acmISBN{978-1-4503-7962-5/20/05}

\begin{document}

\title{Towards Bridging the Gap between\\Control and Self-Adaptive System Properties}

\author{Javier C\'amara}
\affiliation{ \institution{University of York, UK}}
\email{javier.camaramoreno@york.ac.uk}
\author{Alessandro V. Papadopoulos}
\affiliation{ \institution{M\"alardalen University, Sweden}}
\email{alessandro.papadopoulos@mdh.se}
\author{Thomas Vogel}
\affiliation{ \institution{Humboldt University Berlin, Germany}}
\email{thomas.vogel@cs.hu-berlin.de}
\author{Danny Weyns}
\affiliation{ \institution{\mbox{KU Leuven, Belgium; Linnaeus, Sweden}}}
\email{danny.weyns@kuleuven.be}
\author{David Garlan}
\affiliation{ \institution{Carnegie Mellon University, USA}}
\email{garlan@cs.cmu.edu}
\author{Shihong Huang}
\affiliation{ \institution{Florida Atlantic University, USA}}
\email{shihong@fau.edu}
\author{Kenji Tei}
\affiliation{ \institution{Waseda University, Japan}}
\email{ktei@aoni.waseda.jp}
 
\renewcommand{\shortauthors}{C\'amara et al.}

\begin{abstract}
Two of the main paradigms used to build adaptive software employ different types of properties to capture relevant aspects of the system's run-time behavior. On the one hand, control systems consider properties that concern static aspects like stability, as well as dynamic properties that capture the transient evolution of variables such as settling time. On the other hand, self-adaptive systems consider mostly non-functional properties that capture concerns such as performance, reliability, and cost. In general, it is not easy to reconcile these two types of properties or identify under which conditions they constitute a good fit to provide run-time guarantees. There is a need of identifying the key properties in the areas of control and self-adaptation, as well as of characterizing and mapping them to better understand how they relate and possibly complement each other. In this paper, we take a first step to tackle this problem by: (1)~identifying a set of key properties in control theory, (2)~illustrating the formalization of some of these properties employing temporal logic languages commonly used to engineer self-adaptive software systems, and (3)~illustrating how to map key properties that characterize self-adaptive software systems into control properties, leveraging their formalization in temporal logics. We illustrate the different steps of the mapping on an exemplar case in the cloud computing domain and conclude with identifying open challenges in the area.
 \end{abstract}


\maketitle

\section{Introduction}

Two of the main paradigms used to build adaptive software employ different types of properties to capture relevant aspects of the system's run-time behavior. On the one hand, control systems consider properties that concern static aspects like stability, as well as dynamic properties that capture transient aspects such as settling time. On the other hand, self-adaptive systems consider mostly non-functional properties that include concerns such as performance, cost, and reliability. 

Self-adaptive software can clearly benefit from the potential that control theory provides in terms of enabling better analyzability and enforcement of constraints on run-time system behavior.
Being able to formally reason about the non-functional concerns of a system (e.g., security, energy, performance) in terms of control properties in the presence of an unpredictable environment can optimize operation and improve the level of assurances that engineers can provide about the systems they build.

However, applying control theory to software systems poses a set of challenges that do not exist in other domains~\cite{Filieri:2017,TSE.2017.2704579}. 
One of the main challenges is that control-based solutions demand the availability of precise mathematical models that capture both the dynamics of the system under control, as well as the properties that engineers want to impose and reason about.
When control is applied to physical plants, the laws that govern the system are captured by accurate mathematical models that are well-understood, and relevant properties like stability or performance are formally characterized by definitions that are precise and standard in the control community~\cite{astrom2010feedback}.

While obtaining accurate models of non-functional aspects of software behavior can to some extent be achieved using different methods like {\em system identification}~\cite{6213241}, the self-adaptive software systems community still lacks a standard repertoire of run-time properties formally characterized in a way that makes them amenable to formal analysis using techniques applied by software engineers in self-adaptive systems (e.g., run-time verification, model checking). Having such a repertoire would not only help individual system designers express and check certain common fundamental properties, but also help promote norms for system assurance across the community of adaptive systems developers.

Solving in software the kind of problems that control theory solves in other domains entails understanding how control properties relate to software requirements and formally characterizing such properties in a way that facilitates their instantiation and automated analysis using standard tools.

To advance the understanding of how self-adaptive system requirements relate to control properties, in this paper we: (1)~identify a set of key properties in control theory, (2)~illustrate the formalization of some of these properties employing temporal logic languages commonly used to engineer self-adaptive software systems, and (3)~indicate how to map key properties that characterize self-adaptive software systems into control properties, leveraging their formalization in temporal logics. We illustrate the different steps of the mapping on an exemplar case in the cloud computing domain and conclude with identifying open challenges in the area.

\section{Background}
\label{sec:background}

In this section, we first present a basic set of concepts in control systems, followed by a description of a general class of discrete abstractions which are employed to capture the non-functional behavior of self-adaptive systems at run time.

\subsection{Control Terminology}
In this paper, we focus mainly on continuous-time signals and systems, but equivalent definitions are present in the case of discrete-time~\cite{astrom2010feedback}.

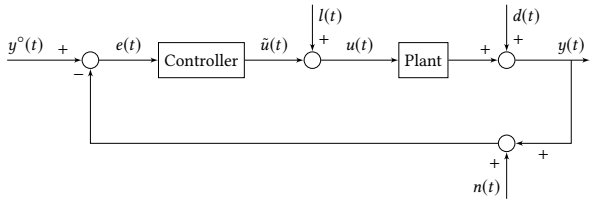
\begin{figure}[h]
\centering
\resizebox{0.95\columnwidth}{!}{
\tikzset{external/figure name={controlSynthesisScheme}}
\begin{tikzpicture}[auto, node distance=1.5cm,>=latex']
 \node[input] (u) {};
 \node[sum,right of=u] (sum) {};
 \node[block,right of=sum,node distance=2cm] (R) {Controller};
 \node[sum,right of=R,node distance=2cm] (sumld) {};
 \node[block,right of=sumld,node distance=2cm] (G) {Plant};
 \node[sum,right of=G] (sumd) {};
 \node[output,right of=sumd]	 (y) {};
 \node[sum,below of=sumd] (sumn) {};
 \node[coordinate,above of=sumld,node distance=1cm] (ld) {};
 \node[coordinate,above of=sumd,node distance=1cm] (d) {};
 \node[coordinate,below of=sumn,node distance=1cm] (n) {};
 
 \draw[->]	(u) -- node[near start]{$y^{\circ}(t)$} node[near end]{$+$} (sum);
 \draw[->]	(sum) -- node{$e(t)$} (R);
 \draw[->]	(R) -- node{$\tilde u(t)$}(sumld);
 \draw[->]  (sumld) -- node{$u(t)$} (G);
 \draw[->]	(G) -- node[near end]{$+$} (sumd);
 \draw[->]	(sumd) --	node[near end](yn){$y(t)$} (y);
 \draw[->]	(yn) |- node[near end]{$+$} (sumn);
 \draw[->]	(sumn) -| node[pos=0.95]{$-$} (sum);
 \draw[->]	(ld)-- node[near start]{$l(t)$} node[near end]{$+$} (sumld);
 \draw[->]	(d) -- node[near start]{$d(t)$} node[near end]{$+$} (sumd);
 \draw[->]	(n) -- node[near start]{$n(t)$} node[near end]{$+$} (sumn);
 
\end{tikzpicture}
}
\caption{Control scheme.}
\label{fig:controlSynthesisScheme}
\end{figure}

First, consider the control scheme represented in Figure~\ref{fig:controlSynthesisScheme}. The two main blocks represent the \textbf{Controller} and the \textbf{Plant} respectively. The Plant is the object that we want to control. Let $t\in \mathbb{R}$ be the continuous-time, where $\mathbb{R}$ indicates the real numbers; all the signals that are introduced next are functions of the time $t$. The $m$ \textbf{inputs} of the plant are represented as $u(t) \in \mathbb{R}^m$, and in computing systems are typically referred as \emph{control parameters}, or \emph{tuning parameters}. The $p$ \textbf{outputs} of the plant are typically represented as $y(t) \in \mathbb{R}^p$, and in computing systems are typically referred as \emph{measurements} or \emph{sensors data}.

For every output $y(t)$ of the plant, one defines a desired behavior for it, which in control terms is called a \textbf{setpoint} or \textbf{reference signal}, and is represented by $y^{\circ}(t) \in \mathbb{R}^p$.

The difference between the desired behavior and the actual behavior of the plant is called \textbf{error}, and is represented as $e(t) \in \mathbb{R}^p$:
\centerline{
$ e(t) = y^{\circ}(t) - y(t).$
}

The controller is a decision-making mechanism that given the error, decides what is the value of the \textbf{control signal} $\tilde{u}(t) \in \mathbb{R}^m$ in order to make the error converge to zero. In principle, the control signal and the plant input should be the same, i.e., $\tilde{u}(t) = u(t)$, but in practice, there might be a \textbf{load disturbance} $l(t) \in \mathbb{R}^m$, that affects the controller decision. Therefore, it holds that\\
\centerline{
$u(t) = \tilde{u}(t) + l(t).$
}
The load disturbance is one of the main disturbances that affect the performance of control systems.

In addition, there might be a disturbance that is acting directly on the output of the plant, which is called \textbf{output disturbance}, and it is represented as $d(t) \in \mathbb{R}^p$. Finally, there is  \textbf{noise} $n(t) \in \mathbb{R}^p$ that affects the measurements that one takes of the output. These two last sources of disturbances are typically ``high-frequency'' disturbances, and can be counteracted by a suitable filtering at design time of the controller.

As a main reference to these concepts, the interested reader can refer to the publicly available book by {\AA}str{\"o}m and Murray~\cite{astrom2010feedback}.

\subsection{Discrete Models}
\label{sec:discretemodel}
We {\it consider the self-adaptive system as a black-box on which a set of output variables can be monitored over time}. Concretely, we model the non-functional run-time behavior of a self-adaptive system as a transition system that captures the evolution over time of a set of relevant variables (i.e., state is characterized by a collection of $n$ real-valued random variables $Y=\{y_1, \ldots, y_n\}$).  
{\it These variables can be considered to be analogous to the outputs $y(t)$ in a control system}.  
Sampling these variables in space and time results in their quantization and time discretization. 

Let $[\alpha_i, \beta_i]$ be the range of $y_i$, with $\alpha_i, \beta_i \in \R$, and $\eta_i \in \R^+$ be its quantization parameter. Then, $y_i$ takes its values in the set: 
\begin{center}
$ [\R]_{y_i}=\left\{r : \R \,\,\left| \; r=k \eta_i ,\; k \in\mathbb{Z} , \; \alpha_i \leq r \leq \beta_i \right.\right\}$.
\end{center}

Hence, given an observed value of $y_i$ at time $t$ (denoted as $y_i(t))$, the corresponding quantized value is obtained as:

\centerline{$quant(y_i(t))= \min (\displaystyle \argmin_{r \in [\R]_{y_i}   }(|y_i(t) - r|)$).}
Variables in $Y$ define a state-space $ [\R^n]_{Y}=[\R]_{y_1} \times \ldots \times [\R]_{y_n}$. 
Furthermore, we assume a time discretization parameter $\tau \in \R^+$ associated with the sampling period established for the observation of variables, determining the transition time.

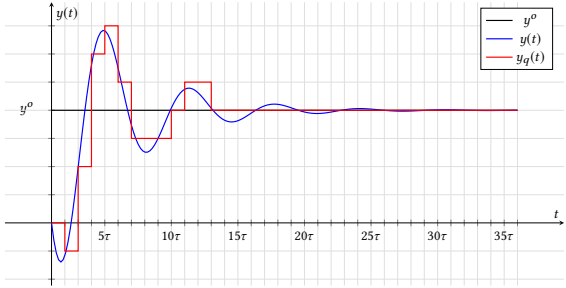
\begin{figure}[h]
\centering
\resizebox{0.9\columnwidth}{!}{

\pgfplotsset{grid style={solid,gray!25}}

\begin{tikzpicture}

\node (inf) at (0.5,4) {$y^o$};


\begin{axis}[
grid=major,
axis lines=middle,
width=0.8\textwidth,
height=8cm,
xtick={0,1,...,35},
xticklabels={,,,,$5\tau$,,,,,$10\tau$,,,,,$15\tau$,,,,,$20\tau$,,,,,$25\tau$,,,,,$30\tau$,,,,,$35\tau$},
ytick={-1,-0.75,...,2},
yticklabels=\empty,
enlargelimits=true,
xlabel={$t$},
ylabel={$y(t)$},
domain=0:35,
legend entries={$y^o$, $y(t)$, $y_q(t)$}
]


\addplot[color=black,line width=0.75pt] {1};

\addplot[color=blue,samples=200,line width=0.75pt] {1 - (exp(-x/5)*(sqrt(6)*cos(deg((2*sqrt(6)*x)/5)) + 3*sin(deg((2*sqrt(6)*x)/5))))/sqrt(6)};


\addplot[const plot, color=red,line width=0.75pt] table[row sep=crcr] {%
0	0\\
1	-0.25\\
2	0.5\\
3	1.5\\
4	1.75\\
5	1.25\\
6	0.75\\
7	0.75\\
8	0.75\\
9	1\\
10	1.25\\
11	1.25\\
12	1\\
13	1\\
14	1\\
15	1\\
16	1\\
17	1\\
18	1\\
19	1\\
20	1\\
21	1\\
22	1\\
23	1\\
24	1\\
25	1\\
26	1\\
27	1\\
28	1\\
29	1\\
30	1\\
31	1\\
32	1\\
33	1\\
34	1\\
35	1\\
};

%
%
%
%
\end{axis}  
\end{tikzpicture}
}
\caption{Discrete quantized vs continuous output.}
\label{fig:discretemodel}
\end{figure}

Figure~\ref{fig:discretemodel} compares an arbitrary continuous system output $y(t)$ with its quantized counterpart $y_q(t)$\footnote{For convenience, we write in the following $y_q(t)$ instead of $quant(y(t))$.} in the discrete timeline.
$y_q(t)$ takes values only in multiples of $\eta_y$, and is represented in the figure as constant for intervals of duration $\tau$.

Discrete models can be enriched with rewards and costs that help capture quantitative aspects of system behavior (e.g., elapsed time, energy consumption, cost) in a precise manner. 
These rewards can be employed as building blocks to reason about properties that capture quantitative aspects of system behavior over time.

A {\it reward structure} is a pair $(\rho,\iota)$, where $\iota :  [\R^n]_{Y} \rightarrow \mathds{R}_{\geq 0}$ is a function that assigns rewards to system states, and $\rho:  [\R^n]_{Y} \times  [\R^n]_{Y} \rightarrow \mathds{R}_{\geq 0}$ is a function assigning rewards to transitions.

State reward $\iota(s)$ is acquired in state $s \in  [\R^n]_{Y}$ per time step, that is, each time that the system spends one time step in $s$, the reward accrues $\iota(s)$. In contrast, $\rho(s,s')$ is the reward acquired every time that a transition between $s$ and $s'$ occurs.

For illustration purposes, we assume that rewards are defined as sets of pairs  $(pd, r)$, where $pd$ is a predicate over states $[\R^n]_{Y}$, and $r \in \mathds{R}_{\geq 0}$ is the accrued reward when $s \in  [\R^n]_{Y} \models pd$. 
If the pair  $(pd, r)$ corresponds to a transition reward, the reward is accrued when a transition from a source state $s \in  [\R^n]_{Y} \models pd$ occurs.

\section{Illustration Exemplar: {RUBiS}}
\label{sec:example}

\begin{figure}[!htbp]
    \centering
    \psfrag{c0}{$c_0$} \includegraphics[width=0.8\linewidth]{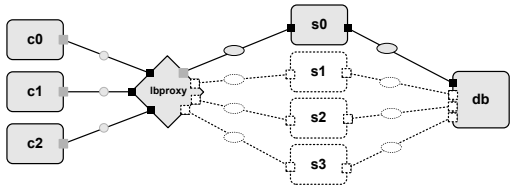}    
    \caption{RUBiS architecture.}
    \label{fig:rubis-arch}
\end{figure}

We illustrate our formalization of properties on RUBiS~\cite{rubis}, an open-source application that implements the functionality of an auctions website. 
Figure~\ref{fig:rubis-arch} depicts the architecture of RUBiS, which consists of a web server tier that receives requests from clients using browsers, and a database tier that acts as a data provider for the web tier. 
The system also includes a load balancer to distribute requests among web servers using a round-robin policy.
When a web server receives a page request from the load balancer, it accesses the database to obtain the data required to render dynamic page content.
The only relevant property of the operating environment that we consider in this scenario is the request arrival rate prescribed by the workload induced on the system. 

The system includes two actuation points that can be operationalized by a controller to make the system self-adaptive and deal with the changing request arrival rate:

\noindent $\bullet$ {\em Server Addition/Removal}. Server addition has an associated latency, whereas the latency for server removal is assumed to be negligible.

\noindent $\bullet$ {\em Dimmer}. The version of RUBiS used for our comparison follows the {\em brownout} paradigm~\cite{DBLP:conf/icse/KleinMAH14}, in which the response to a request includes mandatory content (e.g., the details of a product), and optional content such as recommendations of related products. 
A {\em dimmer} parameter (taking values in the interval $[0,1]$) can be set to control the proportion of responses that include optional content.
The goals of the target system are summarized in two functional and three
non-functional requirements (Table~\ref{tab:reqs}). 

\begin{table}[!h]
\centering
{\footnotesize
\caption{Requirements for RUBiS.}
\label{tab:reqs}
\setlength\tabcolsep{1pt}
{
\begin{tabular}{lp{7.8cm}}
\hline
\multicolumn{2}{c}{\bf Functional Requirements} \\
\hline
{\bf R1} & The target system shall respond to every request for serving its content. \\
{\bf R2}& The target system shall serve optional content to the connected clients.  \\
\hline
\multicolumn{2}{c}{\bf Non-Functional Requirements} \\
\hline
{\bf NFR1}& The target system shall demonstrate high performance. The average
response time $r$ should not exceed $T$. \\

{\bf NFR2}& The target system shall provide high availability of the optional content. 
Subject to NFR1, the percentage of requests with optional content (i.e., the dimmer 
value $d$) should be maximized. \\

{\bf NFR3}& The target operating system shall operate under low cost. Subject to NFR1 
and NFR2, the cost (i.e., the number of servers $s$) should be minimized. \\
\hline
\end{tabular}
}
}
\nnskip
\end{table}

There is a strict preference order among the non-functional requirements that deal with 
optimization, so trade-offs among different dimensions to be optimized are not 
possible (i.e., no solution should compromise maximizing the percentage 
of requests with optional content to reduce cost). 
The imposition of a preference order is aimed at better capturing real scenarios and is not a limitation imposed by any of the compared approaches, which are also able to capture non-strict preference orders among requirements.

\section{Characterizing Control Properties}
\label{sec:ctprops}

Control systems are usually concerned about four main objectives~\cite{Filieri:2017}, namely: (a)~{\em setpoint tracking}, which is related to achieving the specified setpoint whenever it is reachable, (b)~{\em transient behavior}, concerned about how setpoints are reached, in particular in the presence of abrupt changes, (c)~{\em robustness to inaccurate or delayed measurements}, related to the ability of a controller to behave correctly even when transient errors or delayed data is provided to it, and (d)~ {\em disturbance rejection}, related to the ability of avoiding any effect of external interferences on system goals. 
These high level objectives can be mapped in control theory into the satisfaction {\em by design} of properties like {\em stability}, {\em guaranteed settling time}, {\em integrated squared error}, that relate to the achievable runtime performance of the control system. 
In this section, we describe these properties, going from their mathematical formulation into their characterization in temporal logics commonly used in formal verification like LTL~\cite{DBLP:conf/focs/Pnueli77}, CTL~\cite{DBLP:conf/lop/ClarkeE81}, and PCTL~\cite{DBLP:journals/fac/HanssonJ94}. Other properties exist in control theory, but having a complete catalogue here is beyond the scope of this paper, and it is left as future work.

\subsection{Stability}

The concept of \textit{stability} in control theory differs from the notion of stability used in self-adaptive software. A control system is stable even if the error $e(t)$ is not converging to zero, but it is bounded. More specifically, in control terms, if the initial value of system output $y(0)$ is ``close'' to the equilibrium value $y^{\circ}$, then the evolution over time of the output $y(t)$ will be bounded (and not diverge) from $y^{\circ}$. More formally:
\begin{equation}
stby \equiv \forall \epsilon>0 \; \exists\delta(\epsilon) \; | \; \| y(0) - y^{\circ} \|<\delta(\epsilon) \Rightarrow \| y(t) - y^{\circ} \|<\epsilon, \forall t > 0  \label{eqn:stability}
\end{equation}
A system is \textit{asymptotically stable}, if it is stable (as per the previous definition), and also if the evolution over time of the system output will eventually converge to $y^{\circ}$. More formally:
\begin{equation}
    as\_stby \equiv stby \land \lim_{t \to \infty} \| y(t) - y^{\circ} \| = 0
\end{equation}

\begin{figure}[h]
\centering
\resizebox{0.9\columnwidth}{!}{
\tikzset{external/figure name={stepResponseDynamic}}
\begin{tikzpicture}
\begin{axis}[
axis lines=middle,
width=0.8\textwidth,
height=8cm,
xtick=\empty,
ytick=\empty,
enlargelimits=true,
xlabel={$t$},
ylabel={$y(t)$},
domain=0:35,
axis on top,
legend entries={$y^{\circ}(t)$,$y(t)$}
]
\draw [gray,fill=gray!50, opacity=0.5] (axis cs:0,1.05) rectangle (axis cs:35,0.95);

\addplot[color=black] {1};
\addplot[color=blue,samples=200] {1 - (exp(-x/5)*(sqrt(6)*cos(deg((2*sqrt(6)*x)/5)) + 3*sin(deg((2*sqrt(6)*x)/5))))/sqrt(6)};

%
%
%
\draw[dotted] (axis cs:17.0391,1.0521) -- (axis cs:17.0391,0);
\draw[<->] (axis cs:17.0391,0.2) -- node[above]{$t_{s,\epsilon}$} (axis cs:0,0.2);

\draw[<->] (axis cs:36,0.95) -- node[right]{$2\epsilon$} (axis cs:36,1.05);
\end{axis}  
\end{tikzpicture}
}
\caption{Example of system output stabilization.}
\label{fig:stability}
\end{figure}
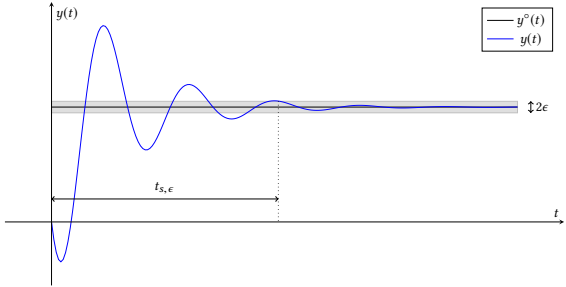
Figure~\ref{fig:stability} shows the response of a system that eventually stabilizes within an error band (gray box) of width $2\epsilon$. 

\medskip
\noindent{\bf Characterization in Temporal Logic}. Characterizing stability in temporal logic requires capturing the constraints imposed by the definition of stability given in Expression~\ref{eqn:stability}. 
Such characterization can be given on a quantized version of the variables and constants required to define stability: 
\begin{equation}
[stby] \equiv \| y_q - y^{\circ}_q \| < \delta_q \Rightarrow \Box (\| y_q - y^{\circ}_q \| < \epsilon_q)
\label{tl:stability}
\end{equation}
In Expression~\ref{tl:stability}, the subscript $q$ indicates that the constant or variable on which it appears is the quantized version of its continuous counterpart (i.e., $y_q(t) \equiv quant(y(t))$, cf. Section~\ref{sec:discretemodel}). 
It is worth noticing that variables in a software system are quantized by definition, and requiring a notion of asymptotic stability may be too restrictive. The current definition captures the same concept with $\epsilon_q$ being the resolution of the quantization or a tolerance parameter.
Moreover, the absence of explicit time indexes is consistent with the implicit notion of time introduced by the temporal operators. 
For instance, when $y_q$ is not within the scope of any temporal operator (like in the antecedent of the implication given in the formula), the expression refers to the value of the variable in the first state of the trace (i.e.,  $y_q \equiv y_q(0)$). 
However, if the same term is within the scope of a temporal operator as it happens with the $\Box$ on the right-hand side of the expression, then the same $y_q$ refers to the value of $y_q(t)$ in all subsequent states of the discrete temporal line (i.e., $y_q(t)$ when $t=0, \tau, 2\tau, \dots$). The non-probabilistic version of this property is directly expressible in LTL and CTL (as ${\sf \bf A}[stby]$), whereas its probabilistic version can employ the probability quantifier of PCTL (e.g., $P_{=?}[stby]$, $P_{\leq b}[stby]$).

\noindent {\bf Instantiation in RUBiS.} Expression~\ref{tl:stabilityrub} instantiates $[stby]$, in a straightforward manner for the response time variable $r$, assuming a setpoint equivalent to the threshold $T$. It states that when the error becomes smaller than $\delta^r_q$ it will stay within the band $[T-\epsilon^r_q,T+\epsilon^r_q]$.
\begin{equation}
 \| r_q - T \| < \delta^r_q \Rightarrow \Box (\| r_q - T \| < \epsilon^r_q)
\label{tl:stabilityrub}
\end{equation}

\subsection{Settling Time}
\label{sec:settlingtime}
One of the key indicators of how the system reaches its goals is \textit{settling time} $t_s$, which is the time needed by the system to reach a new steady-state equilibrium.

For an arbitrary $\epsilon \in \mathds{R}^+$, the \textit{$\epsilon$-settling time} is defined by:
\begin{equation}
t_{s,\epsilon} \equiv \inf \{\delta \; {\text s.t.} \; \|y(t) - y^{\circ}\| < \epsilon, \forall t \in [\delta, \infty]\}
\label{eqn:settlingtime}
\end{equation}

In Expression~\ref{eqn:settlingtime}, the settling time is captured as the infimum of the set of time values in the continuous timeline for which the error is bounded by $\epsilon$ in the following. 
Note that the infimum is the greatest lowest bound that always exists, meaning that it takes the value $\infty$ if the stability condition is never satisfied.

\noindent{\bf Characterization in Temporal Logic}. In contrast with stability, which is a boolean property that is either satisfied by the system or not (Expression~\ref{tl:stability}), settling time is a quantitative property and therefore we characterize it as a temporal logic expression that employs a reward quantifier. 
Since in this case the reward captures time, we assume the existence of a transition reward function ${\sf [time]} \equiv (true, \tau)$ that accrues the time quantum employed for time in the discrete model whenever a transition in the discrete timeline is taken:
\begin{equation}
[t_{s,\epsilon}] \equiv {\sf R^{[time]}_{=?}}[\Diamond \Box  \| y_q - y^{\circ}_q \| < \epsilon_q]
\label{tl:settlingtime}
\end{equation}
Expression~\ref{tl:settlingtime} characterizes the settling time as the time reward accrued until the system reaches a state from which the error is bounded by $\epsilon_q$. 
There are two aspects of this characterization that are important to highlight. 
First, the reachability formula accrues reward until it reaches a state that satisfies the reachability predicate, but the reward in the latter state is not included. 
Second, when the reachability predicate is not satisfied, the semantics of the reward quantifier assign an infinite reward as the value that is obtained when the expression is quantified (e.g., in PCTL, co-safe LTL with rewards).  
These two aspects make this characterization consistent with the definition given in Expression~\ref{eqn:settlingtime}, which defines the settling time as the time instant immediately prior to the one in which the error is already bound by $\epsilon$, and becomes infinite if the error is not always bound by $\epsilon$, starting at some arbitrary point in the timeline. 
Note that, due to the nesting of temporal operators including $\Box$, this property is not (currently) directly expressible in temporal logics with Markovian rewards as implemented in probabilistic model checkers like PRISM~\cite{DBLP:conf/cav/KwiatkowskaNP11} and Storm~\cite{DBLP:conf/cav/DehnertJK017}. However, assuming finite traces in our discrete models, we can perform a preprocessing step on the traces, labeling explicitly states from which $\Box \| y_q - y^{\circ}_q \| < \epsilon_q$ as $p$, and then model check the property as:
\begin{equation}
{\sf R^{[time]}_{=?}}[\Diamond p] \label{tl:simpler}
\end{equation}

\noindent {\bf Instantiation in RUBiS.} Expression~\ref{tl:settlingtimerubis} instantiates $[t_{s,\epsilon}] $ with similar assumptions to those adopted for Expression~\ref{tl:stabilityrub}.
\begin{equation}
{\sf R^{[time]}_{=?}}[\Diamond \Box  \| r_q - T \| < \epsilon^r_q]
\label{tl:settlingtimerubis}
\end{equation}

\subsection{Integrated Squared Error}
\label{sec:ise}
\begin{figure}[!h]
\centering
\resizebox{0.9\columnwidth}{!}{
\tikzset{external/figure name={stepResponseDynamic}}
\pgfmathdeclarefunction{poly}{0}{\pgfmathparse{1 - (exp(-x/5)*(sqrt(6)*cos(deg((2*sqrt(6)*x)/5)) + 3*sin(deg((2*sqrt(6)*x)/5))))/sqrt(6)}}

\begin{tikzpicture}
%
%
%
%

\begin{axis}[
axis lines=middle,
width=0.8\textwidth,
height=8cm,
xtick=\empty,
ytick={1},
yticklabels={$y^{\circ}$},
enlargelimits=true,
xlabel={$t$},
ylabel={$y(t)$},
domain=0:35,
axis on top,
legend entries={$y^{\circ}(t)$,$y(t)$}
]
\draw [gray,fill=gray!50, opacity=0.5] (axis cs:0,1.05) rectangle (axis cs:35,0.95);

\addplot[name path=line, gray, no markers, line width=1pt] {1};
\addplot[name path=poly, black, thick, mark=none, samples=200] {poly};

\addplot fill between[ 
    of = poly and line, 
    split, 
    every even segment/.style = {red!25},
    every odd segment/.style  = {blue!25}
  ];

%
%
%
\draw[dotted] (axis cs:17.0391,1.0521) -- (axis cs:17.0391,0);
\draw[<->] (axis cs:17.0391,0.2) -- node[above]{$t_{s,\epsilon}$} (axis cs:0,0.2);

\draw[<->] (axis cs:36,0.95) -- node[right]{$2\epsilon$} (axis cs:36,1.05);
\end{axis}  
\end{tikzpicture}
}
\caption{Illustration of the integrated squared error.}
\label{fig:ise}
\end{figure}
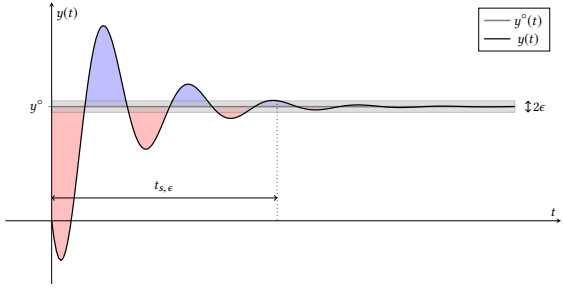

Relevant quantitative measures of a system's performance are also often based on the behavior of the error $e(t)$. 
We consider here as a representative index the {\em integrated squared of the error} (ISE):
\begin{equation}
\text{ISE} \equiv \int_{0}^T e^2(t) \mathrm{d}t
\label{eqn:ise}
\end{equation}
The ISE integrates the square of the error over time (see Figure~\ref{fig:ise}), penalizing large errors more than smaller ones (the square of a large error will be much bigger). 
Control systems specified to minimize ISE of the tracking error $e(t)$, e.g., MPC or LQG~\cite{camacho2004model}, tend to eliminate large errors quickly, but tolerate small ones persisting for a long period of time. This often leads to fast responses, but with considerably low-amplitude oscillation.

\medskip
\noindent{\bf Characterization in Temporal Logic}. Similar to the settling time, the ISE is a quantitative property and therefore we characterize it making use of a reward quantifier. 
Since in this case the reward has to capture accrued error over time, we assume the existence of a transition reward function ${\sf [error]} \equiv (true, (\| y_q - y^{\circ}_q\|)^2)$ that accrues the square of the instantaneous error whenever a transition in the discrete temporal line is taken. 

Then, we can write an expression that accrues the error reward over the discrete timeline before stability is achieved:
\begin{equation}
[ISE]  \equiv {\sf R^{[error]}_{=?}}[\Diamond \Box  \| y_q - y^{\circ}_q \| < \epsilon_q]
\label{tl:ise}
\end{equation}
Due to the nesting of $\Diamond \Box$, this property is not directly expressible in PCTL/Co-safe LTL with rewards. However, under the same assumptions described for the settling time property, a similar model preprocessing step can enable its practical verification through a simpler probabilistic reachability property (cf. Expression~\ref{tl:simpler}).

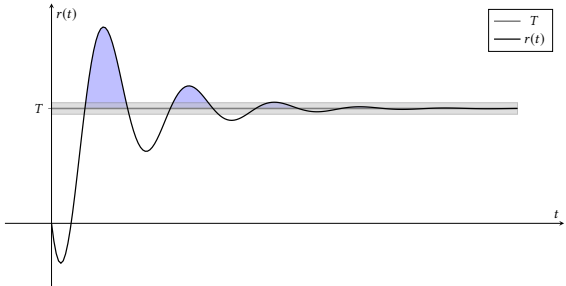
\begin{figure}[h]
\centering
\resizebox{0.9\columnwidth}{!}{

\pgfmathdeclarefunction{poly}{0}{\pgfmathparse{1 - (exp(-x/5)*(sqrt(6)*cos(deg((2*sqrt(6)*x)/5)) + 3*sin(deg((2*sqrt(6)*x)/5))))/sqrt(6)}}

\begin{tikzpicture}


%
%
%

\begin{axis}[
axis lines=middle,
width=0.8\textwidth,
height=8cm,
xtick=\empty,
ytick={1},
yticklabels={$T$},
enlargelimits=true,
xlabel={$t$},
ylabel={$r(t)$},
domain=0:35,
axis on top,
legend entries={$T$, $r(t)$}
]

\draw [gray,fill=gray!50, opacity=0.5] (axis cs:0,1.05) rectangle (axis cs:35,0.95);

%

\addplot[name path=line, gray, no markers, line width=1pt] {1};
\addplot[name path=poly, black, thick, mark=none, samples=200] {poly};

\addplot fill between[ 
    of = poly and line, 
    split, 
    every even segment/.style = {white!0},
    every odd segment/.style  = {blue!25}
  ];

%
%
%
%
\end{axis}  
\end{tikzpicture}
}
\caption{Example of RUBiS performance response with accrued positive squared error.}
\label{fig:iserubis}
\end{figure}

\noindent {\bf Instantiation in RUBiS}. We assume that RUBiS is working on steady state, but suddenly receives a spike on request arrival rate, causing the average response time $r$ to go above threshold $T$ (Figure~\ref{fig:iserubis}). 
After violating the threshold, the system adds a server to drive down the response time below $T$. 
Before the system stabilizes, its response time may experience some oscillations that make $r$ go above and below $T$ several times. 
For simplicity, we assume $y^{\circ}=T$.

To obtain an indication of how well the system is adapting, we can employ a modified version of the $[ISE]$ property (Expression~\ref{tl:ise}). 
In this case, we are only interested in accruing a penalty whenever the output of the system is above the threshold $T$, therefore we adapt the reward structure for the error, constraining it to accrue reward only whenever $r>T$, i.e., ${\sf [penalty]} \equiv (r>T, (r-T)^2)$: 
\begin{equation}
{\sf R^{[penalty]}_{=?}}[\Diamond \Box  \| r-T \| < \epsilon_q] \label{tl:iserubis}
\end{equation}
We can observe that the accrued error corresponds to the colored areas enclosed by $T$ and $r(t)$ in Figure~\ref{fig:iserubis}. 
Since negative error (i.e., when $r<T$) does not constitute a violation of the response time threshold, we do not accrue it, in contrast with the more general property described in Expressions~\ref{eqn:ise} and~\ref{tl:ise}.

\section{Formalizing Non-Functional Requirements}

The non-functional run-time behavior of self-adaptive systems can be captured by an external observer as a set of quantitative indicators that represent attributes of different concerns such as performance, cost, or availability. In this section, we employ the characterization of control properties in temporal logic introduced in the previous section as building blocks to formalize non-functional requirements in RUBiS.

\noindent {\bf NFR1}. {\em The target system shall demonstrate high performance. The average response time $r$ should not exceed $T$.} This requirement can be captured by combining temporal logic properties of: (i)~stability as described by Expression~\ref{tl:stabilityrub}, (ii)~settling time as captured by Expression~\ref{tl:settlingtimerubis}, and (iii)~an integrated error property analogous to Expression~\ref{tl:iserubis} using the penalty,  ${\sf [penalty]} \equiv (r>T, r-T)$  which should be guaranteed to be always less or equal to zero, i.e.:
\begin{equation}
{\sf R^{[penalty]}_{=?}}[\Diamond \Box  \| r-y^{\circ}_r \| < \epsilon_q] \leq 0 \label{tl:iserubisnfr1}
\end{equation}
Note that in the expression above, the error term $\| r-y^{\circ}_r \|$ does not make the simplifying assumption included in Expression~\ref{tl:iserubis}, and incorporates an arbitrary setpoint different from $T$. This makes sense in a realistic setting because, if $y^{\circ}_r=T$, oscillations around the setpoint during transients would always result in response time threshold violations. This is also applicable to properties (i) and (ii) for this requirement. 

\noindent {\bf NFR2}.{\em The target system shall provide high availability of the optional content. Subject to NFR1, the percentage of requests with optional content (i.e., the dimmer value $d$) should be maximized.} Capturing this requirement requires instantiating the integrated error property on variable $d$, which should be always as close as possible to 1 (maximum optional content):
\begin{equation}
{\sf R^{[optional]}_{=?}}[\Diamond \Box  \| 1-d_q  \| < \epsilon^ d_q]
\label{tl:ised}
\end{equation}
where ${\sf [optional]} \equiv (true, d)$. Note that in this case, the magnitude of the error is always below 1, so minimizing the non-squared error is a more sensible choice.

 \noindent {\bf NFR3}. {\em The target operating system shall operate under low cost. Subject to NFR1 and NFR2, the cost (i.e., the number of servers $s$) should be minimized.} The formalization of this requirement can be captured using the following properties defined over the response time variable $r$: (i)~stability as described by Expression~\ref{tl:stabilityrub}, (ii)~settling time as captured by Expression~\ref{tl:settlingtimerubis}. Finally, we can capture the penalty of using extra servers during the transient by employing an integral error property which should minimize the use of servers according to ${\sf [penalty]} \equiv (r>T, s^2)$:
 \begin{equation}
 {\sf R^{[penalty]}_{=?}}[\Diamond \Box  \| r_q-T  \| < \epsilon^ r_q]
 \label{tl:ised}
 \end{equation}
 Note that in this case, stability and settling time properties are defined over response time $r$, whereas penalty is defined over the number of servers employed $s$, making an interesting case in which formalizing a single requirement involves combining different control properties across variables.

All variables might present similar patterns in terms of control properties, but the composition of the self-adaptive properties is non-trivial and might be realized in different ways. As a consequence, there is a need to incorporate high-level compositional operators to enable joint evaluation of the requirements. Alternatively, we might want to express priorities in how specific properties should be achieved.

\section{Related Work}
\label{sec:related}

We have grouped related work in four parts: control applied to computing systems, automatically generated control solutions, verification of control properties, and evaluation of quality properties.  

\vspace{2pt}\noindent\textbf{Control Applied to Computing System.} In 2004, Hellerstein et al. wrote a pioneering book on applying control theory to computing systems~\cite{Hellerstein:2004}. 
Over the years, control-based approaches have been applied extensively to computing systems, mostly focussing on controlling lower-level resources. Abdelzaher et al. apply different types of controller models (e.g., PI and PID) to deal with performance requirements of servers~\cite{1200252}. Wang et al. present DEUCON that allocates local controllers to computing units that only coordinate with neighbors~\cite{4218578}. Stability analysis is based on the location of poles of the composite system's transfer function. Imes et al. present CoPPer, a control-theoretic approach that applies adaptive control to meet soft performance goals by manipulating hardware power limits~\cite{8831193}. In contrast, our work targets a mapping between classic control properties and typical software qualities.

\vspace{2pt}\noindent\textbf{Automatically Generated Control Solutions.}
To deal with the complexity of control theory, researchers have started investigating automatic generation of control solutions to adapt software~\cite{Filieri:2017, ShevtsovWM19,Weyns:2018}. 
Filieri et al. introduce the push-button methodology (PBM) that automatically constructs a linear model of a software system for a PI controller to adapt the system for one setpoint goal~\cite{Filieri:2012}. 
Shevtsov et al. propose a solution to control a software system for multiple goals, including an optimization goal~\cite{Shevtsov:2016}. 
Maggio et al. apply model-predictive control (MPC) to software adaptation~\cite{Maggio:2017}, while Anagelopoulos et al. apply a requirements-driven approach with MPC~\cite{Angelopoulos:2018}. These approaches highlight properties that are important from a control-theoretic viewpoint, but this accounts for only one side of the problem we target in this paper, namely, a rigorous specification and verification of classic control properties.  

\vspace{2pt}\noindent\textbf{Verification of Control Properties.} Some work exits on the formalisation and verification of properties of control systems. We highlight two representative examples. 
Preuse and Hanisch apply model checking to verify safety, liveliness and deadlock properties of manufacturing control systems that are specified in temporal logic~\cite{5970316}. 
Yan et al. use approximate bisimulation for comparing the similarity between a complex (continuous) cyber-physical system and a (discretized) higher level model of it~\cite{YanJLWZ16}. The authors illustrate the approach for a safety property. Our work complements these approaches by focusing on typical software quality properties and the formal mapping of these with control properties. 

\vspace{2pt}\noindent\textbf{Evaluation of Quality Properties.}
A number of approaches zoom in on the evaluation of quality properties in self-adaptive systems. 
Weyns and Ahmad~\cite{Weyns:2013} performed a systematic literature review identifying the  main quality properties considered in self-adaptation: efficiency/performance of the system (55\% of the studies), reliability (41\%), and flexibility (28\%). 
Reinecke et al.~\cite{Reinecke:2010} propose a payoff metric to measure the ``success'' of adaptation. This metric is a user-defined function aggregating QoS metrics observed on the running system similarly to a utility function. 
Villegas et al.~\cite{Villegas:2011} present a framework to evaluate adaptation properties, i.e., stability, accuracy, settling time, overshoot, robustness, termination of adaptation, consistency, scalability, and security. 
The properties are informally defined and mapped to software qualities based on examples from literature. 
Raibulet et al.~\cite{Raibulet:2017} focus on quality attributes to evaluate the utility of a self-adaptive system, and software metrics to evaluate the quality of the adaptation at runtime, whereas C\'amara and de Lemos~\cite{DBLP:conf/icse/CamaraL12} evaluate resilience properties formalized in PCTL. 
Each of these approaches contributes to a better understanding of quality properties and their evaluation from a software engineering point of view. However, this only accounts for one side of the mapping problem we target in this paper, i.e., a traditional software engineering perspective. 

\vspace{2pt}\noindent\textbf{Conclusion.}
While control theory and self-adaptive systems contribute knowledge about properties in their domain, there is little understanding on the mapping between the two types of properties, which is precisely the target of the research presented in this paper. 

\section{Conclusions and Future Work}
\label{sec:roadmap}

In this paper, we have taken the first step in bridging the gap between control and self-adaptive system properties.
We have (1)~identified key properties in control theory (stability, settling time, and integral error), 
(2)~formalized these properties in temporal logic languages, which are typically used to specify properties (requirements) of software systems,
and (3)~illustrated how non-functional properties of self-adaptive systems (performance, availability, and costs) can be mapped into these control properties by using this formalization and the RUBiS exemplar.
To achieve the formalization and mapping, we have discussed the abstraction of transition systems describing discrete state spaces on which self-adaptive system attributes are measured and how this abstraction is able to represent continuous system dynamics in which control properties are typically characterized.
Models of such transition systems and control properties formalized in a temporal logic can serve as input for off-the-shelf run-time verification tools and model checkers.

This approach advances the understanding of how non-functional requirements relate to control properties (e.g., which requirements can be characterized by which control properties) and paves the way for an improved operation and assurance of self-adaptive systems via formal reasoning (e.g., by run-time verification) based on control. 
Our approach is currently limited by the set of control properties that we have formalized, requirements that we have mapped into control properties (cf. previous paragraph), and the expressiveness of temporal logics, which might not be able to fully capture the nuances of some control properties (cf. Section~\ref{sec:settlingtime}).

Our long-term goal is to understand whether control theory can be used as a formal foundation for specifying and analyzing self-adaptive systems, and if so, under which conditions. Towards that goal, work is needed to identify further corresponding and complementing properties between self-adaptive and control systems (e.g., whether real-time or security requirements can be mapped into control properties), and to leverage the formalization of properties for a formal assessment of controllers in self-adaptive systems (e.g., to provide guarantees for the correctness of controllers). 

\balance
\bibliographystyle{ACM-Reference-Format}
\bibliography{biblio}

\end{document}